# Influence of the magnetic filter field topology on the beam divergence at the ELISE test facility

M. Barbisan[1, a)], U. Fantz[2] and D. Wünderlich[2]

[1] *Consorzio RFX (CNR, ENEA, INFN, Univ. of Padua, Acciaierie Venete SpA),*
*C.so Stati Uniti 4 – 35127, Padova (Italy).*
[2] *Max-Planck-Institut für Plasmaphysik (IPP), Boltzmannstr. 2, D-85748 Garching, Germany.*

[a)]Corresponding author: marco.barbisan@igi.cnr.it

**Abstract.** The ELISE test facility hosts a RF negative ion source, equipped with an extraction system which should deliver half the current foreseen for the ITER Neutral Beam Injector, keeping the ratio of co-extracted electrons to ions below 1. An important tool for the suppression of the co-extracted electrons is the magnetic filter field, produced by a current flowing in the plasma grid, the first grid of the 3 stage extraction system. To boost the source performances new concepts for the production of the magnetic filter field have been tested, combining the existing system with permanent magnets attached on the source walls. The topologies of these new magnetic configurations influence the beam particles' trajectories in the extraction region, with consequences for the overall beam optics. These effects will be characterized in this article by studying the angular distribution of the beam particles, as measured by the Beam Emission Spectroscopy (BES) diagnostic. The behavior of the beam will be studied also through the measurements of the currents flowing on the grounded grid (the third grid) and on the grid holder box surrounding its exit. The main finding is that the broader component of the beam increases when the magnetic field is strengthened by permanent magnets, i.e. in the cases in which most of the co-extracted electrons are suppressed.

## INTRODUCTION

The ELISE test facility represents an important step towards the development of the Neutral Beam Injectors (NBIs) for the ITER reactor. ELISE [1-5] hosts a $H^-/D^-$ ion source composed by 4 RF plasma drivers, coupled to an expansion chamber in which Cs is evaporated to increment the $H^-/D^-$ production. The expansion chamber is coupled to the so called Plasma Grid (PG), from which negative ions are extracted and subsequently accelerated. The whole acceleration system comprehends:

- The Plasma Grid itself;
- The Extraction Grid (EG), with embedded magnets to dump the electrons co-extracted from the source;
- The Grounded Grid (GG);
- The Grounded Grid Holder Box and the Green Shield (GGGHB), a structure which surrounds the GG exit and acts as an electrostatic shield for the beam. A 3D representation of the GGGHB is given in Fig. 1.

ELISE was designed to deliver the same beam current density (286 A/m$^2$ for D, 329 A/m$^2$ for H) foreseen for the ITER NBIs, from an extraction area (0.1 m$^2$) which is half the one of the future ITER NBIs. Cesium is evaporated in the source in order to boost the production of negative ions on the PG surface. The plasma in the source can be sustained for up to 1 hour, while the beam extraction can be repeatedly performed for 10 s at intervals of ≈150 s.

Besides reaching high levels of extracted current and extraction duration, another target of the experimentation on ELISE is keeping the fraction of co-extracted electrons below 1. This is accomplished by means of these 2 methods: firstly, the PG is biased positively with respect to the source body and to a Bias Plate (BP), which surrounds the PG apertures upstream the acceleration system. Secondly, a magnetic filter field is created in the expansion chamber, in proximity of the PG; the magnetic field reduces the electron flux on the PG but also lowers the electron temperature, reducing the negative ion destruction rate determined by electron impact. In the standard configuration the magnetic filter field is produced by a current $I_{PG}$ (max. 5 kA) which flows vertically through the PG. As example, for a current of 2.5 kA a field of 2.4 mT is generated at 66 mm upstream the center of the PG. The magnetic field can be also strengthened or weakened (±0.4 mT at 66 mm upstream the center of the PG) by adding external permanent magnets on the lateral walls of the expansion chamber. The 3 possible magnetic configurations (no external magnets, external magnets strengthening or weakening the standard field) don't simply give a different magnetic field intensity, but the whole topology of the field in space is different [6-7]. The magnets give rise to a more intense magnetic field in their proximity (i.e. on the source lateral walls). Moreover, while the magnetic field produced by the magnets has the same polarity upstream and downstream the PG, the polarity of the field generated by the PG current flips at the axial center of the PG surface. As consequence, the magnetic field produced by $I_{PG}$ can strengthened in the source by the magnets but weakened in the acceleration system, and vice versa.

The magnetic field strength and topology used at ELISE directly affect the amounts of ions and electrons extracted, but also the trajectories of the beam particles in the extraction region (i.e. between PG and EG) and in the acceleration region (i.e. between EG and GG). Aim of this paper is to show how the properties of the produced beam are affected by the topologies of the magnetic filter field. In particular, the angular distribution of the beam particles under different magnetic field intensities and topologies will be shown and discussed. Studying this is important because in future neutral beam injectors a poor focalization of the resulting beam could cause undesired heat loads to beam line components. The information on the beam angular distribution has been obtained by the analysis of the data collected by the Beam Emission Spectroscopy (BES) diagnostic [8]. The BES data are supported also by the measurement of the currents flowing in EG, GG and GGGHB; indeed, these components can be hit by electrons and negative ions sufficiently deviated from the axes of the grids apertures, giving information about the beam along its path.

The paper firstly illustrates the features of the BES diagnostic and of the method used to analyze the experimental data. Results of electric measurements on the above-mentioned components are presented and then correlated with the data from the BES diagnostic.

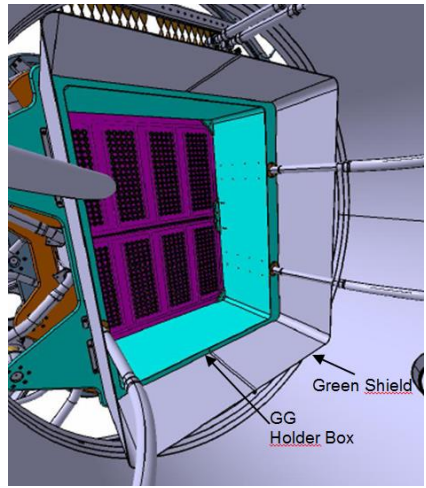

**Figure 1.** CAD drawing of the Grounded Grid Holder Box and the Green Shield, looking upstream towards the GG (depicted in purple).

## THE BEAM EMISSION SPECTROSPY DIAGNOSTIC

Beam Emission Spectroscopy is a noninvasive tool that allows to measure divergence and uniformity of the beam, plus the fraction of beam particles neutralized by stripping reactions in the acceleration system. The BES diagnostic is based on the spectral analysis of the $H_\alpha/D_\alpha$ radiation emitted by the beam [9]. The $H_\alpha/D_\alpha$ photons come from hydrogen atoms excited by collisions between the beam particles and the $H_2/D_2$ molecules of the residual gas in the vacuum tank. The photons, collected by several optic heads along Lines of Sight (LOSs), have a wavelength λ' which is Doppler shifted with respect to the nominal $H_\alpha/D_\alpha$ wavelength ($\lambda_0$=656.2793 nm for H, $\lambda_0$=656.1032 nm for D), according to the formula

$$\lambda' = \lambda_0 \frac{1 - \beta \cos\alpha}{\sqrt{1 - \beta^2}} \qquad (1)$$

where β is the ratio between the emitting particle's speed and the speed of light, while α is the angle between the emitting particle's trajectory and the direction of the received photon. An example of the comprehensive spectrum of the $H_\alpha/D_\alpha$ light collected by a LOS is shown in Fig. 2(a); the unshifted peak, at nominal $\lambda_0$ wavelength, is related to photons produced by slow atoms created by collisions of the beam with the background gas involved in the collisions: their speed is too low to cause a detectable Doppler shift. Part of the unshifted peak is due also to stray light generated in the source and passing through the grids apertures. The other 2 peaks shown in the figure are related to the Doppler shifted photons coming from the fast excited atoms present in the beam. The most intense peak, the one at 660 nm in the example, is related to beam particles accelerated to full energy, given by the voltage difference between PG and GG and then neutralized. The other, less intense peak is instead given by beam particles which were only partially accelerated and neutralized by stripping losses inside the acceleration system. From the ratio of the intensities of the 2 peaks it is possible to calculate the stripping loss fraction. From the linewidth of the full energy Doppler peak it is possible to retrieve the beam divergence ε; the angular distribution of the beam particles indeed determines the distribution of α, and then of λ'. If the sigma width $\sigma_D$ (half width at $1/e^{0.5}$ maximum) of the full energy Doppler peak is known, for example by means of a Gaussian fit, then it is possible to calculate the e-folding (half width at 1/e maximum) beam divergence as

$$\varepsilon = \sqrt{2} \cdot \sqrt{\frac{\sigma_D^2 - \sigma_S^2}{[(\lambda' - \lambda_0)\tan\alpha]^2} - \omega^2} \qquad (2)$$

where $\sigma_S$ is the sigma width of the spectrometer instrumental function and ω represents a fluctuation in the collection angle due to the finite dimensions and the focusing of the optic heads. In the case of ELISE, $\sigma_S$=17.9 pm, ω=0.16°, while α, given by the orientation of the optic heads, is equal to 130°.

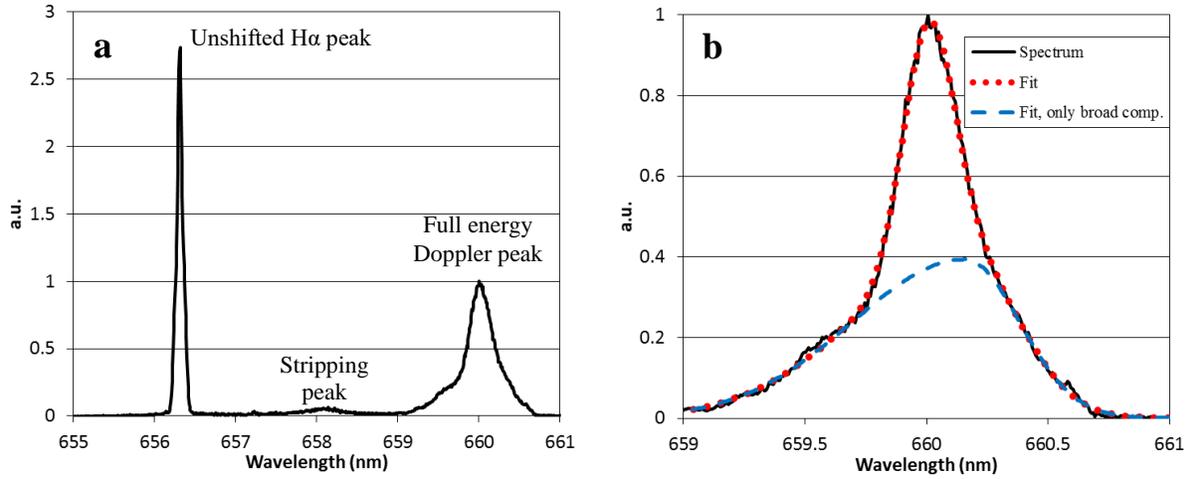

**Figure 2.** Plot (a): example of a spectrum collected by the BES diagnostic in ELISE, during a pulse in hydrogen. Plot (b): example of fit of the full energy Doppler peak with the sum of a Gaussian plus an asymmetric bi-Gaussian). The black curve represents the measured spectrum, the same of plot (a), while the red dotted curve indicates the whole fitted function. The blue dashed curve indicates the only bi-Gaussian, representing the broad component of the peak.

The angular distribution of the beam particles can be complex, so that the full energy Doppler peak cannot be simply fitted with a Gaussian shape. The peak can be thought as the sum of an often asymmetric broad component, visible at the peak base and corresponding to an angularly spread component of the beam, and a narrow component, roughly corresponding to the half top of the peak. This model, whose validity will be further discussed in [10], has been translated into a fitting function to get information about the beam. This function is the sum of an asymmetric bi-Gaussian, representing the beam broad component, plus a Gaussian for the narrow component. More precisely, the asymmetric bi-Gaussian is the junction of two halves Gaussian curves with different widths but same center and height. An example of the fit of the full energy Doppler peak with this function is shown in Fig. 2(b): the Doppler peak measured by BES is indicated with a black curve, while the fitted curve is represented with a red dotted curve. Additionally indicated with a blue dashed curve is the only bi-Gaussian.

From the fit of the Doppler peak with this function it is possible to get: the fraction of the broad component integral over the whole peak integral, the divergence related to the narrow component and the average divergence of the broad component, calculated as the average of the divergence values obtained by the 2 linewidths of the bi-Gaussian.

In this paper the properties of the broad component will be discussed, since the most spread particles of the beam can represent a concern, in terms of thermal loads, for the beam line components of future neutral beam injectors.

Regarding the measurements of the fraction of broad component, the values given by the analysis of the spectra represent an underestimation (between -10% and -30%) of the broad component of the beam. Indeed, from ad-hoc simulations of the beam and of the BES diagnostic [11] it has been understood that the fraction of beam volume intersected by a LOS is lower when the divergence is higher. The broad and narrow components of the Doppler peak then derive from different volumes of the LOS, so that the fraction of broad component is underestimated. In spite of this, the relations of majority/minority between measurements of this type remain valid.

The measurements given by the BES diagnostic in ELISE are available for 16 horizontal LOSs, spaced 5 cm vertically, and 4 vertical LOS, spaced 16 cm horizontally. The LOSs intercept the beam axis at an average distance of 2.75 m from the GG. The 20 optic heads which collect the light have a clear aperture of 20 mm and are connected via optical fibers to an *Acton Spectra Pro 750i* spectrometer, coupled to a CCD camera of 1024x1024 pixels, 13 µm wide. The spectra can be acquired with a plate factor of 7 pm/pixel.

## EXPERIMENTAL RESULTS

To characterize the influence of the magnetic filter field on the beam properties, a specific set of pulses executed in ELISE has been selected. The pulses were performed in hydrogen, at 0.6 Pa source pressure, in underperveant conditions and with Cs evaporated in the source. The pulses were carried out at different values of PG current and for all the 3 magnetic configurations (no external magnets, external magnets strengthening or weakening the standard field).

The impact of modifying the magnetic field topology on the extracted currents is shown in Fig. 3: in plot (a) the EG current $I_{EG}$, mainly attributable to co-extracted electrons deviated by the EG magnets, is shown as function of $I_{PG}$; in plot (b) the extracted current $I_{ion}$ (i.e. the beam current flowing downstream the EG) is plotted against $I_{PG}$. In these plots and in all the following ones the data taken with the standard field configuration are represented with red squares, while those acquired with permanent magnets strengthening or weakening the field in the source are indicated with blue diamonds and green triangles, respectively.

What shown in Fig. 3(a) indicates that, as expected, the current of co-extracted electrons (in first approximation equal to $I_{EG}$) decreases with increasing magnetic field in the source, because of both $I_{PG}$ and the chosen magnetic configuration. In particular, $I_{EG}$ is significantly high at low values of $I_{PG}$ and with the configuration weakening the field in the source; in this case $I_{EG}$ can even exceed the extracted negative ion current (see Fig. 3(b)).

Regarding $I_{ion}$, Fig. 3(b) shows that the extracted current, as expected, slightly decreases when the magnetic filter field in the source is increased by $I_{PG}$ or by the magnetic configuration. It must be noticed that for the "weakened field" configuration $I_{ion}$ does not steadily decrease with $I_{PG}$, but has a maximum for a certain value of $I_{PG}$. What observed in Fig. 3(b) for the "weakened field" configuration could be explained by a modification of the plasma structure determined by the magnetic field topology. Another possible explanation, complementary to the first, is that the magnetic field topology influences the trajectory of Cs ions and then the distribution of Cs on the PG surface [12].

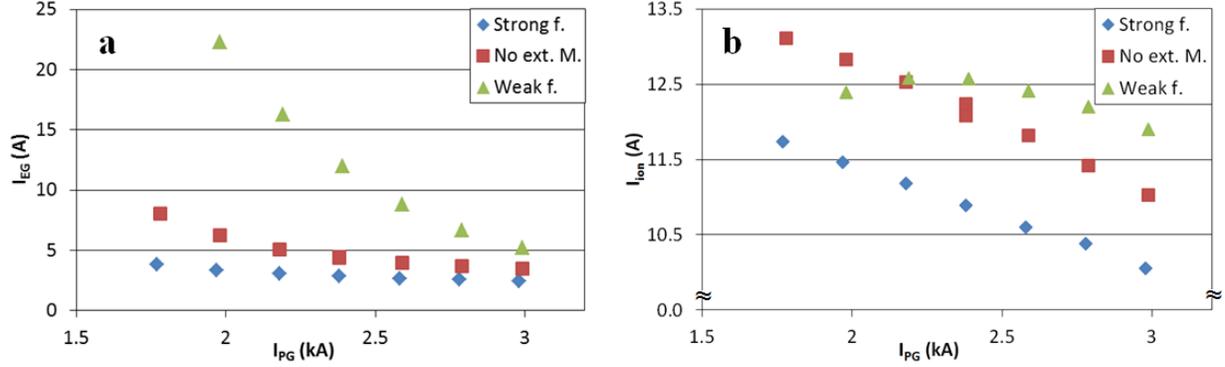

**Figure 3.** Measurements of EG current ($I_{EG}$-plot (a)) and extracted current ($I_{ion}$-plot (b)), plotted as function of the PG current ($I_{PG}$). The considered pulses were performed with different values of $I_{PG}$ and for different magnetic configurations of the source: with permanent magnets strengthening the magnetic field produced by $I_{PG}$ in the source (blue diamonds), with no magnets (red squares) and with magnets weakening the field in the source (green triangles).

According to the BES analysis of the above discussed pulses, the average divergence of the broad component of the beam results to be between 6.8° and 7.5° e-folding, independently on the magnetic configuration. No dependence on $I_{PG}$ was observed, at least within the statistical fluctuations of the measured values.

The fraction of the broad component composing the full energy Doppler peak resulted to increase with the intensity of the magnetic field in the source, varied by acting on $I_{PG}$ or the magnetic configuration. This is shown in Fig. 4(a), where the average fraction of broad component, measured from the LOSs intercepting the most intense regions of the beam, is plotted as function of $I_{PG}$. The trend of the points related to the "weakened field" configurations showed a local minimum, contrarily to what observed in the other 2 configurations. This behavior may be an effect of the magnetic filter field topology on the plasma and on the Cs conditioning at the PG.

The fraction of the broad component of the Doppler peak was also compared with the amount of extracted beam particles (electrons/ions) which, being too divergent, have impacted on the GG. The result of this comparison is is shown in Fig. 4(b) where the fraction of broad component is plotted against the ratio between the GG current $I_{GG}$ and the extracted beam current. In general, it is found that the fraction of broad component always increases together with the losses on the GG. The increase of the amount of spread beam particles with the magnetic filter field is then confirmed both in the acceleration region and beyond the GG. It must be noticed that, given a certain fraction of broad component, the losses on the GG are higher if the magnets are weakening the field in the source. The particular trend of the points of this configuration is just due to the specific dependency of the fraction of broad component on $I_{PG}$ (Fig. 4(a)).

At last, the information on the broad component of the Doppler peak is compared to the current collected on the GGGHB. This current is caused by the impact on the GGGHB of two types of charged particles: one are the electrons released by the neutralization reactions undergone by the beam ions. The other one is represented by beam particles (eg. the beam halo) whose trajectories are strongly deviated from the beam axis. It was calculated that a negative ion exiting from one of the most lateral holes of the GG should travel with an angle ≥7.7° on the horizontal plane to hit the GGGHB. The value of this angle is compatible with the average divergence attributed to the broad component, therefore variations of the intensity of the broad component can lead to variations in the GGGHB current $I_{GGGHB}$. This is shown in Fig. 4(c), where the fraction of the broad component is plotted against the fraction of the accelerated beam current (i.e. $I_{ion}$-$I_{GG}$) which has been collected on the GGGHB. The data indicate that an increase of the fraction of broad component, because of $I_{PG}$ or of the magnetic configuration, leads to an increase of the fraction of accelerated current collected on the GGGHB. This confirms the relation between the broad component and the current collected on the GGGHB. The only exception is given by the data related to the "weakened field" configuration: at low $I_{PG}$ the fraction of broad component varies with negligible changes on the GGGHB current.

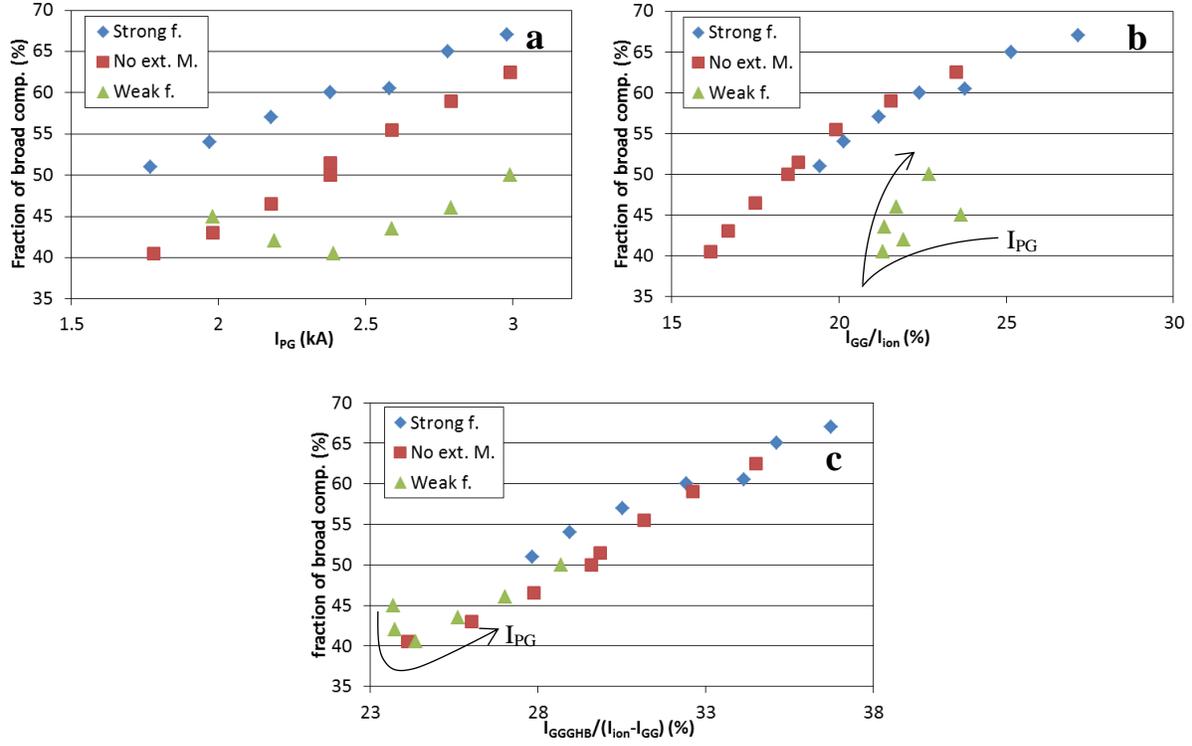

**Figure 4.** Average values of the fraction of broad component of the Doppler peak, plotted as function of: the PG current ($I_{PG}$), the fraction of the extracted current ($I_{ion}$) collected on the GG and the fraction of the accelerated current ($I_{ion}$-$I_{GG}$) collected on the GGGHB. The considered pulses (the same of Fig. 4) were performed with different values of $I_{PG}$ and for different magnetic configurations of the source: with permanent magnets strengthening the magnetic field produced by $I_{PG}$ in the source (blue diamonds), with no magnets (red squares) and with magnets weakening the field in the source (green triangles). For a comparison with plot (a), in plot (b) and plot (c) the trend of $I_{PG}$ has been indicated for the points of the "weakened field" configuration.

## CONCLUSIONS

From electrically measured currents and from information given by the BES diagnostic at ELISE it is possible to estimate how the magnetic filter field influences the properties of the generated beam. As expected, a higher intensity of the magnetic field in the source leads to a lower amount of co-extracted electrons, at the price of a minor reduction in extracted negative ions.

However, increasing the magnetic field, by raising $I_{PG}$ or by changing the filter field configuration, additionally leads to an increase of the fraction of broad component in the beam. The modifications of the magnetic field cannot be expressed just in terms of intensity, but the whole topology of the field changes from one magnetic configuration to another. This is what probably makes the "weakened field" configuration so different from the other two; the amount of electrons collected on the EG and the losses on the GG make this configuration unsuitable for the optimization of source and acceleration system in ELISE.

The magnetic filter field intensity in the source of future Neutral Beam Injectors will have to be adjusted to a compromise level: not too low to avoid massive co-extraction of electrons, but also not too high to avoid significant damages on the beam line components, because of the power carried by the broad component of the beam. The experimentation on the ELISE test facility will advance to find the optimum magnetic topology for the production of negative ions; moreover, it will be further investigated how the magnetic field topology is able to affect the plasma and the Cs conditioning.

## ACKNOWLEDGMENTS

The work was supported by a contract from Fusion for Energy (F4E-2009-0PE-32-01), represented by Antonio Masiello. The opinions expressed herein are those of the authors only and do not represent the Fusion for Energy's official position.

## REFERENCES

1. P. Franzen, U. Fantz, W. Kraus, H. Falter, B. Heinemann, R. Nocentini and the NNBI Team, *Physical and Experimental Background of the Design of the ELISE Test Facility*, AIP Conf. Proc. **1097**, 451 (2009).
2. B. Heinemann, H. Falter, U. Fantz, P. Franzen, M. Fröschle, R. Gutser, W.Kraus, R. Nocentini, R. Riedl, E. Speth, A. Stäbler, D. Wünderlich, P. Agostinetti, T. Jiang, *Design of the "half-size" ITER neutral beam source for the test facility ELISE*, Fus. Eng. Des. **84** (2009), pp. 915–922.
3. U. Fantz, P. Franzen, B. Heinemann, and D. Wünderlich, *First results of the ITER-relevant negative ion beam test facility ELISE*, Rev. Sci. Instrum. 85, 02B305 (2014).
4. P. Franzen, D. Wünderlich, R. Riedl, R. Nocentini, F. Bonomo, U. Fantz, M. Fröschle, B. Heinemann, C. Martens, W. Kraus, A. Pimazzoni, B. Ruf and NNBI Team, *Status of the ELISE test facility*, AIP Conf. Proc. **1655**, 060001 (2015).
5. U. Fantz, P. Franzen, W. Kraus, L. Schiesko, C. Wimmer and D. Wünderlich, *Size scaling of negative hydrogen ion sources for fusion*, AIP Conf. Proc. **1655**, 040001 (2015).
6. W. Kraus, U. Fantz, B. Heinemann and D. Wünderlich, *Concepts of magnetic filter fields in powerful negative ion sources for fusion*, Rev. Sci. Instrum. **87**, 02B315 (2016) and Rev. Sci. Instrum. **87**, 059901 (2016).
7. D. Wünderlich, W. Kraus, M. Fröschle et al, paper submitted to Plasma Phys. Control. Fusion.
8. R. Nocentini, U. Fantz, P. Franzen, M. Fröschle, B. Heinemann, R. Riedl, B. Ruf, D. Wünderlich and the NNBI team, *Beam diagnostic tools for the negative hydrogen ion source test facility ELISE*, Fus. Eng. Des. **88** (2013), pp. 913– 917.
9. P. Franzen, U. Fantz, and the NNBI Team, *Beam Homogeneity Dependence on the Magnetic Filter Field at the IPP Test Facility MANITU*, AIP Conf. Proc. **1390**, 310 (2011);
10. M. Barbisan, U. Fantz and D. Wünderlich, *Beam characterization by means of emission spectroscopy in the ELISE test facility*, paper in course of submission.
11. M. Barbisan, *Beam emission spectroscopy studies in a H-/D- beam injector*, PhD Thesis, Università degli Studi di Padova (2015).
12. D. Wünderlich, U. Fantz, B. Heinemann, W. Kraus, R. Riedl and the NNBI team, *Long pulse, High Power Operation of the ELISE Test Facility*, MonO3 contribution to NIBS'16 conference.